\documentclass[12pt,a4wide]{article}
\usepackage{afterpage}
\usepackage{epsfig}
\usepackage[a4paper]{geometry}
\usepackage{float}
\usepackage[stable]{footmisc}
\usepackage[utf8x]{inputenc}
\usepackage{a4wide}
\usepackage{amsmath,amssymb,amstext,amsthm,amssymb}
\usepackage{amsfonts}
\usepackage{mathrsfs}
\usepackage{framed}
\usepackage{graphicx}
\usepackage{bbold}
\usepackage{slashed} 
\usepackage{tensor} 
\usepackage{braket} 
\usepackage{array}
\usepackage{graphicx}
\usepackage{setspace}
\usepackage{hyperref}
\usepackage{overpic}
\usepackage{bbm}
\usepackage{cite}
\usepackage{color}
\usepackage{lipsum}
\usepackage{bm}

\DeclareSymbolFontAlphabet{\mathbb}{AMSb}

\let\baraccent=\=
\renewcommand{\=}[1]{\stackrel{#1}{=}}

\begin{document}

\pagestyle{plain}

\makeatletter
\@addtoreset{equation}{section}
\makeatother
\renewcommand{\theequation}{\thesection.\arabic{equation}}
\pagestyle{empty}

\vspace{0.5cm}

\begin{center}
{\LARGE \bf{A Cosmological Constant}\\
\bf{That is Too Small}
		\\[15mm]}
\end{center}

\begin{center}
\scalebox{0.95}[0.95]{{\fontsize{14}{30}\selectfont Mehmet Demirtas,$^{a,b}$ Manki Kim,$^{a,c}$ Liam McAllister,$^{a}$}} \vspace{0.35cm}
\scalebox{0.95}[0.95]{{\fontsize{14}{30}\selectfont Jakob Moritz,$^{a}$ and Andres Rios-Tascon$^{a}$}}
\end{center}

\begin{center}
\vspace{0.25 cm}
\textsl{$^{a}$Department of Physics, Cornell University, Ithaca, NY 14853, USA}\\
\textsl{$^{b}$Department of Physics, Northeastern University, Boston, MA 02115, USA}\\
\textsl{$^{c}$Center for Theoretical Physics, MIT, Cambridge, MA 02138 USA}\\

	 \vspace{1cm}
	\normalsize{\bf Abstract} \\[8mm]
\end{center}
\begin{center}
	\begin{minipage}[h]{15.0cm}
	
We construct a vacuum of string theory in which the magnitude of the vacuum energy is
$< 10^{-123}$ in Planck units.
Regrettably, the sign of the vacuum energy is negative, and some supersymmetry remains unbroken.

	\end{minipage}
\end{center}
\newpage
\setcounter{page}{1}
\pagestyle{plain}
\renewcommand{\thefootnote}{\arabic{footnote}}
\setcounter{footnote}{0}
%
%
\setcounter{tocdepth}{2}
%
\newpage
\section{Introduction}

Explaining the smallness of the observed dark energy density \cite{SupernovaSearchTeam:1998fmf,SupernovaCosmologyProject:1998vns},
\begin{equation}
\rho_{\mathrm{obs}} \approx 10^{-123}M_{\mathrm{pl}}^4\,,
\end{equation}
is a profound challenge \cite{Weinberg:1988cp,Polchinski:2006gy,Bousso:2007gp}.  The idea that anthropic selection in a landscape could account for the smallness of $\rho_{\mathrm{obs}}$ \cite{Weinberg:1987dv}
has been influential, and in broad strokes appears to align with known and suspected properties of string theory \cite{Bousso:2000xa}.
Toy landscapes of flux configurations in compactifications of string theory \cite{Ashok:2003gk,Denef:2004ze,Douglas:2006es} admit a vast number $\mathcal{N}_{\mathrm{vac}}$ of solutions,
and the values that the classical vacuum energy $\rho_{\mathrm{class}}$ takes in such ensembles can be finely-spaced, with differences $\Delta \rho_{\mathrm{class}} \sim M_{\mathrm{pl}}^4/\mathcal{N}_{\mathrm{vac}} \ll \rho_{\mathrm{obs}}$.

One natural objection is that the true explanation might be a richer one, involving dynamics or a different sort of mechanism, and we might fail to seek it by too-readily accepting an anthropic approach.  We will do nothing to overcome this concern, remarking only that in the absence of any framework for a dynamical explanation, perhaps the best one can do is to sharpen the landscape argument.

A more practical complaint is that exploring a landscape of vacua of string theory, and finding solutions therein in which the actual value $\rho_{\mathrm{vac}}$  of the vacuum energy obeys $\rho_{\mathrm{vac}} \approx \rho_{\mathrm{obs}}$, appears out of reach.  The mechanism proposed by Bousso and Polchinski \cite{Bousso:2000xa} relies on the high dimensionality of the space of fluxes, and so one faces a search for special, exponentially rare solutions in a high-dimensional energy landscape.  At least in its full generality, this problem's computational complexity makes it inaccessible by a direct assault \cite{Denef:2006ad,Denef:2017cxt,Halverson:2018cio}.  Even worse, given a candidate de Sitter vacuum in string theory,
one can at best imagine computing $\rho_{\mathrm{vac}}$
order by order, and nonperturbatively:
in the string loop expansion, in the $\alpha'$ expansion, and perhaps in other approximations at the same time.  Achieving in this way a precision of order $\rho_{\mathrm{obs}}$ does not seem possible to us.
The difficulty of exhibiting de Sitter vacua of string theory with $\rho_{\mathrm{vac}} \approx \rho_{\mathrm{obs}}$ thus presents an obstruction to bringing the landscape argument for the cosmological constant problem into sharp focus in quantum gravity.

An essential part of the problem is achieving \emph{scale separation} in cosmological solutions of quantum gravity: can one find isolated solutions of string theory in which the
Kaluza-Klein radius $R_{\mathrm{KK}}$
of the internal space is not extremely large in units of the Planck length $\ell_{\mathrm{p}}$, and yet the noncompact spacetime has a
cosmologically-large radius of curvature,
$R_{\mathrm{cosm.}} \sim 10^{60}\,\ell_{\mathrm{p}}$?  In a theory with unbounded continuous parameters this question might not arise.  However, the known classes of realistic solutions of string theory come in large but finite families (see, however, \cite{DeWolfe:2005uu}), and their low-energy parameters are ultimately determined by inherently \emph{quantized} parameters of the string vacuum, such as the topological data of a Calabi-Yau flux compactification.
Whether or not $R_{\mathrm{cosm.}} \gtrsim 10^{60}\,\ell_{\mathrm{p}}$ can occur in a particular setting evidently depends on the ranges of values that these quantized parameters can take, and on how $\rho_{\mathrm{vac}}$ depends on these parameters.

In this note we exhibit a solution of type IIB string theory with $|\rho_{\mathrm{vac}}| \approx 10^{-144}M_{\mathrm{pl}}^4$, in which the internal space --- an orientifold of a Calabi-Yau threefold --- has radius $R_{\mathrm{KK}} \approx 10^{4} \ell_{\mathrm{p}}$.
This vacuum is an example of extreme scale separation, but in $\mathrm{AdS}_4$, not $\mathrm{dS}_4$: the solution preserves $\mathcal{N}=1$ supersymmetry, and the cosmological constant is negative.

Although this compactification cannot describe our universe, it provides an intriguing angle on the cosmological constant problem.
To stumble upon a solution with vacuum energy $\varepsilon M_{\mathrm{pl}}^4$ in a complex landscape of vacua in which the characteristic scale is $M_{\mathrm{pl}}$, one would naively expect to have to explore $\mathcal{O}\bigl(1/\varepsilon\bigr)$
distinct vacua.  The actual distribution of cosmological constants could play a role, of course, but the above expectation should be a reasonable guide unless the set of solutions manifests an exponentially strong concentration around $\rho=0$ --- which is to say, unless the landscape furnishes a bona fide statistical solution to the cosmological constant problem, a possibility that we shall discount in this work.  In particular, to find
vacuum energy $\varepsilon M_{\mathrm{pl}}^4$ in a high-dimensional landscape of flux vacua,
one generally has to search through $\mathcal{O}\bigl(1/\varepsilon\bigr)$ choices of quantized flux.

We have arrived at $|\rho_{\mathrm{vac}}| \ll  10^{-123}M_{\mathrm{pl}}^4$
\emph{without} performing such a costly search.
Instead, we have exploited structures in the quantized parameters occurring in string compactifications: specifically, three-form flux quanta in an orientifold of a Calabi-Yau threefold $X$, and the Gopakumar-Vafa invariants \cite{Gopakumar:1998ii,Gopakumar:1998jq}
of curves in the mirror threefold $\widetilde{X}$.  We have found choices of flux for which the part of the flux superpotential \cite{Gukov:1999ya} that descends from
perturbative contributions to the prepotential in compactification  of type IIA string theory on $\widetilde{X}$ vanishes \emph{exactly}, while the nonperturbative terms, which arise from worldsheet instantons on $\widetilde{X}$, fall into the form of a racetrack.  The competition of two such instanton terms then generates an exponentially small number \cite{Demirtas:2019sip}. When the fluxes and  Gopakumar-Vafa invariants are modest integers like 1 or 2, the resulting hierarchy is similarly modest.  But there exist, and we have found, Calabi-Yau threefolds for which the integers in question are, for example, 2 and 252, and the hierarchy in the vacuum energy is proportional to
\begin{equation}
\left(\frac{2}{252}\right)^{58} \approx 10^{-122}\,.
\end{equation}
In this paper we will present one such example.

\section{A Vacuum}

To define a
compactification\footnote{Extensive discussion of this and other examples can be found in the companion paper \cite{longpaper}: in particular, the full data of the construction is available in the supplemental material associated to the arXiv e-print \cite{longpaper}.} of type IIB string theory on an orientifold of a Calabi-Yau threefold hypersurface $X$ in a toric variety $V$,
we consider the reflexive polytope $\Delta$ \cite{Kreuzer:2000xy} with vertices given by the columns of
\begin{equation}\label{eq:Delta_vertices_main}
\begin{pmatrix}
1  & -3 & -3 & 0  & 0  & 0 &-5 &-2  \\
0 & -2  & -1 & 0  & 0  & 1 &-3 &-1  \\
0 & 0   & -1 & 0  & 1  & 0 & 0 & 1 \\
0 & 0   & 0  & 1  & 0  & 0 &-1 &-1
\end{pmatrix}\, .
\end{equation}
We define  toric varieties $\widetilde{V}$ and $V$, respectively, in terms of fine regular star triangulations of $\Delta$ and its polar dual $\Delta^\circ$.
The anti-canonical hypersurfaces in $\widetilde{V}$ and $V$ are a mirror pair of Calabi-Yau threefolds $\widetilde{X}$ and $X$,
with
$h^{1,1}(X)=h^{2,1}(\widetilde{X})=113$, and $h^{2,1}(X)=h^{1,1}(\widetilde{X})=5$.
We find an O3/O7 orientifold involution of $X$ with 26 O7-planes and 48 O3-planes, and
obeying $h^{1,1}_-=h^{2,1}_+=0$.  The D3-brane tadpole is 60.

In this compactification, the three-form fluxes
\setcounter{MaxMatrixCols}{20}
\begin{align}
&\vec{f}=\begin{pmatrix}
10&  12 & 8 & 0 & 0 & 4 & 0 & 0 & 2 & 4 & 11 & -8
\end{pmatrix}\, ,\nonumber\\
&\vec{h}=\begin{pmatrix}
0 &  8 & -15 & 11 & -2 & 13 & 0 & 0 & 0 & 0 & 0 & 0
\end{pmatrix}\, ,
\end{align}
which carry D3-brane charge 56,
lead to an exponentially small flux superpotential.
The dominant instantons along the perturbatively flat valley,
as defined in \cite{Demirtas:2019sip}, have
Gopakumar-Vafa invariants $-2$ and $252$,  respectively.
The flux superpotential takes the form
\begin{equation}
 W_{\text{flux}}(\tau) = \sqrt{\tfrac{2^3}{\pi^5}}   \Bigl(-2\,e^{2\pi i \tau \cdot \frac{7}{29}}+252\,e^{2\pi i \tau \cdot \frac{7}{28}}\Bigr) M_{\mathrm{pl}}^3+\mathcal{O}\left(e^{2\pi i \tau \cdot \frac{43}{116}}\right)  \, .
\end{equation}
The string coupling is $g_s \approx 0.011$, and we find
\begin{equation}
W_0:= \langle |W_{\text{flux}}|\rangle \approx 0.526 \times \left(\frac{2}{252}\right)^{29} M_{\mathrm{pl}}^3\approx 6.46\times 10^{-62}M_{\mathrm{pl}}^3\,.
\end{equation}
Turning now to the K\"ahler moduli, we find 114 prime toric divisors $D_I$ that are rigid, and whose uplifts to divisors in F-theory have trivial intermediate Jacobian.  These divisors therefore generate nonvanishing superpotential terms \cite{Witten:1996bn}, and their Pfaffian prefactors $\mathcal{A}_{D_I}$ are pure constants \cite{Witten:1996hc}.  Thus the total superpotential takes the form
\begin{equation}\label{kkltwsimple}
  W = W_0 + \sum_{I} \mathcal{A}_{D_I}\,\mathrm{exp}\Bigl(-\tfrac{2\pi}{c_{I}} T_{D_I}\Bigr)+\ldots\,,
\end{equation} where $c_{I}=6$ if there is an $\mathfrak{so}(8)$ stack of seven-branes on $D_I$, and $c_{I}=1$ otherwise.

\begin{figure}
	\centering
	\includegraphics[keepaspectratio,width=17cm]{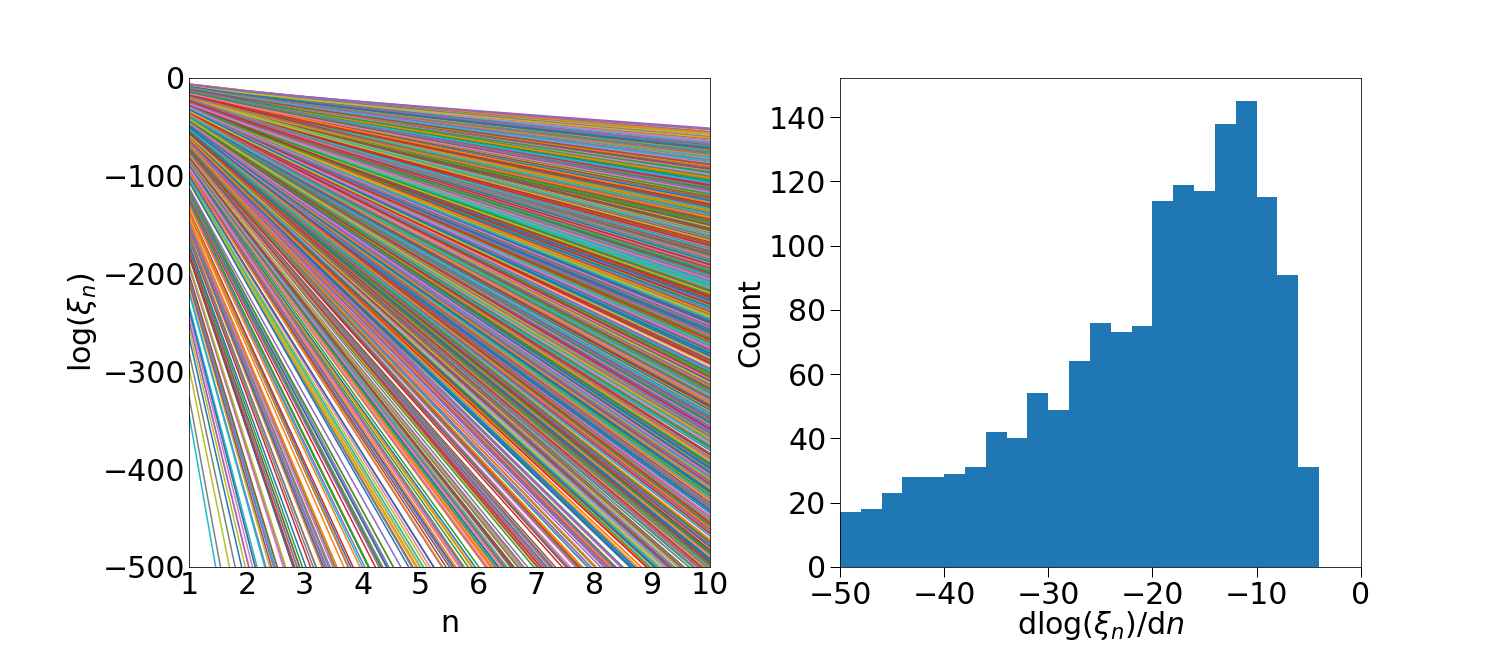}
	\caption{Left: the logarithm of the size $\xi_n$ of the $n$-th term in the series of worldsheet instantons resulting from a sample of 1728 rays in $H_2(X)$. Right: histogram of the slopes of the lines in the left panel.  The sum evidently converges.}
	\label{fig:convergence_5-113-4627}
\end{figure}

With this superpotential we find a $\mathcal{N}=1$ supersymmetric $\mathrm{AdS}_4$ vacuum at which the volume of the internal space in string units is $\mathcal{V} \approx 945$.  At
this point in K\"ahler moduli space  there are perturbative and nonperturbative corrections to the K\"ahler potential $\mathcal{K}$, but these lead to a controllably small \emph{multiplicative} correction to the vacuum energy \cite{longpaper}.
The overall scale of the vacuum energy is dictated by the
superpotential, which does not suffer renormalization at any perturbative order  \cite{PhysRevLett.57.2625,Burgess:2005jx}.
Convergence of the series of worldsheet instanton corrections to $\mathcal{K}$ is shown in Figure \ref{fig:convergence_5-113-4627}.

We have therefore found a supersymmetric AdS$_4$ vacuum with vacuum energy
\begin{equation}
V_0=-3M_{\mathrm{pl}}^{-2}e^{\mathcal{K}}|W|^2\approx  -1.68 \times 10^{-144}M_{\mathrm{pl}}^4\,.
\end{equation}
The Kaluza-Klein radius is $R_{\mathrm{KK}} \approx 10^4\, \ell_p$, so the hierarchy of scales is $R_{\mathrm{cosm.}}/R_{\mathrm{KK}} \approx 10^{68}$.
We have constructed a number of comparable vacua in flux compactifications on orientifolds of Calabi-Yau hypersurfaces with $4 \le h^{2,1} \le 7$ and $51 \le h^{1,1} \le 214$ \cite{longpaper}.

\section{Discussion}

The cosmological constant problem
demands that we
explain
how exponentially large universes arise in theories with small fundamental length-scales.
We have taken a step in this direction by exhibiting an AdS$_4$ solution of string theory in which the AdS length is approximately $10^{72} \ell_{\mathrm{p}}$, but
the Kaluza-Klein radius remains modest, $R_{\mathrm{KK}} \approx 10^4 \ell_p$.

A key feature of our class of constructions is the exact cancellation of all perturbative terms in the superpotential.
Specifically, we have arranged the quantized three-form fluxes so that all perturbative\footnote{The Gukov-Vafa-Witten flux superpotential \cite{Gukov:1999ya} in a type IIB compactification on a Calabi-Yau threefold $X$ is purely classical, but it descends from a prepotential that,
from the viewpoint of type IIA on the mirror $\widetilde{X}$, contains both perturbative and nonperturbative terms.} contributions to the flux superpotential cancel perfectly: this is possible because the superpotential is ultimately determined by integer data.  All remaining terms in the scalar potential are proportional to nonperturbative effects.
We have made further discrete choices to arrange that these effects balance in a controlled minimum, finding an orientifold of a Calabi-Yau threefold $X$ that supports an array of nonperturbative superpotential terms from Euclidean D3-branes and gaugino condensation, and whose mirror $\widetilde{X}$ enjoys a pattern of Gopakumar-Vafa invariants that give rise to a racetrack of worldsheet instantons \cite{Demirtas:2019sip}.  These choices lead to stabilization of all the moduli in a supersymmetric vacuum, as foreseen in \cite{Kachru:2003aw}, with a cosmological constant that is exponentially small
in Planck units.

The smallness of the vacuum energy in our construction is \emph{natural}, in the sense that it is determined by a competition among exponentials, once we have ensured the absence of all perturbative terms in the superpotential by making discrete choices of topology and fluxes.
Of course, it is very well-known that small scales can arise dynamically in supersymmetric theories \cite{Witten:1981nf}, and in four-dimensional $\mathcal{N}=1$ effective supergravity one can easily write a racetrack superpotential whose minimization yields a supersymmetric AdS$_{4}$ vacuum with small vacuum energy.  But presented with such an effective description, one could ask how finely-balanced a racetrack is allowed by ultraviolet completion in quantum gravity.
Thus, something is gained by realizing such a construction
in string theory, where all the underlying parameters are quantized, and their allowable values can be determined.

Our solution is well-controlled as a result of the residual supersymmetry as well as the smallness of the string coupling.
However, it is completely unrealistic, and not solely because the cosmological constant is negative and $\mathcal{N}=1$ supersymmetry is unbroken.
Although the Kaluza-Klein scale is not very far from the Planck scale, some of the moduli are extremely light, with masses proportional to
$W_0/M_{\mathrm{pl}}^2$.  For this reason, we find it implausible that one could construct a realistic model by achieving supersymmetry breaking and an uplift to de Sitter space in the particular compactification we have described here.

We hasten to state
that the fundamental problem of arranging for small positive vacuum energy after supersymmetry breaking has \emph{not} been addressed by this work, and it is not yet clear that our constructions will aid in that quest.  Even so, one can speculate that the smallness of the observed vacuum energy in our universe might be governed by nonperturbative effects in a configuration in which perturbative contributions exactly vanish, and the exactness of the cancellation is ensured by the quantization of fundamental parameters.
In this spirit, exploring solutions of the sort we have presented might lead to a new approach to the cosmological constant problem.

\section*{Acknowledgements}
We thank Yuval Grossman for inspiring discussions,
and we are indebted to
Naomi Gendler, Ben Heidenreich, Tom Rudelius, and Mike Stillman for collaborations on related topics.
The research of M.D., M.K., L.M., and A.R.-T.~was supported in part by NSF grant PHY-1719877, and that of L.M.~and J.M.~was supported in part by the Simons Foundation Origins of the Universe Initiative.
%

\bibliography{refs}
\bibliographystyle{JHEP}
\end{document}